\documentclass[12pt]{article}
\usepackage{amsmath}
\usepackage{amsfonts}
\usepackage{amssymb}

\usepackage{graphics}     
\usepackage{epsfig}
\usepackage{color}        
\usepackage{nicefrac}
\usepackage{mathrsfs}
\usepackage{mathtools}

\newlength{\TZ}
\newcommand{\BEQ}{\begin{equation}}     
\newcommand{\BEA}{\begin{eqnarray}}
\newcommand{\BD}{\begin{displaymath}}
\newcommand{\EEQ}{\end{equation}}       
\newcommand{\EEA}{\end{eqnarray}}
\newcommand{\ED}{\end{displaymath}}
\newcommand{\bb}{\begin{eqnarray}}
\newcommand{\ee}{\end{eqnarray}}
\newcommand{\nn}{\nonumber}             
\newcommand{\e}{{\rm e}}                

\newcommand{\D}{{\rm d}}                
\newcommand{\erfc}{{\rm erfc\,}}        

\newcommand{\demi}{\frac{1}{2}}         
\newcommand{\lap}[1]{\overline{#1}}     

\renewcommand{\vec}[1]{\boldsymbol{#1}} 

\catcode`\@=11
\def\numberbysection{\@addtoreset{equation}{section}
        \def\theequation{\thesection.\arabic{equation}}}
\numberbysection

\definecolor{gruen}{rgb}{0,0.625,0}     
\definecolor{rot}{rgb}{0.75,0,0}        
\definecolor{blau}{rgb}{0,0,0.75}       

\begin{document}

\begin{titlepage}

\vskip 1.5 cm
\begin{center}
{\Large \bf Crossover between diffusion-limited \\[0.15truecm] and reaction-limited regimes in the \\[0.17truecm] coagulation-diffusion process}
\end{center}

\vskip 2.0 cm
\centerline{{\bf Dmytro Shapoval}$^{a,b}$, {\bf Maxym Dudka}$^{a,c}$, {\bf Xavier Durang}$^{d}$ and {\bf Malte Henkel}$^{b,c,e}$}
\vskip 0.5 cm
\begin{center}
$^a$ Institute for Condensed Matter Physics, National Academy of Sciences of Ukraine, \\
1 Svientsitskii Street, UA -- 79011 Lviv, Ukraine
\\ \vspace{0.5cm}
$^b$ Laboratoire de Physique et Chimie Th\'eoriques (CNRS UMR 7019), \\ Universit\'e de Lorraine Nancy,
B.P. 70239,  F -- 54506 Vand{\oe}uvre-l\`es-Nancy Cedex, France
\\ \vspace{0.5cm}
$^c$ ${\mathbb L}^4$ Collaboration \& Doctoral College for the Statistical Physics of Complex Systems,
     Leipzig-Lorraine-Lviv-Coventry, Europe
\\ \vspace{0.5cm}
$^d$ Department of Physics, University of Seoul, Seoul 02504, Republic of Korea
\\ \vspace{0.5cm}
$^e$ Centro de F\'{i}sica Te\'{o}rica e Computacional, Universidade de Lisboa, \\P--1749-016 Lisboa, Portugal
\\ \vspace{1.0cm}
\end{center}

\begin{abstract}
The change {from the diffusion-limited to the reaction-limited cooperative behaviour in reaction-diffusion systems
is analysed by comparing the universal long-time behaviour of the coagulation-diffusion process on a chain and on the Bethe lattice.}
{On a chain}, this model is exactly solvable through the empty-interval method.
{This method can be extended to the Bethe lattice, in the ben-Avraham-Glasser approximation.}
{On the Bethe lattice, the analysis of the Laplace-transformed time-dependent particle-density is analogous to the study of
the stationary state, if a stochastic reset to a configuration of uncorrelated particles is added.
In this stationary state logarithmic corrections to scaling are found, as expected for
systems at the upper critical dimension.}
Analogous results hold true for the time-integrated particle-density.
The crossover scaling functions and the associated effective exponents
{between the chain and the Bethe lattice} are derived.
 \\
\end{abstract}

\vfill
\end{titlepage}

\section{Introduction}
\setcounter{footnote}{0}

Relaxation phenomena far from equilibrium continue to raise important questions in fundamental and applied research.
Diffusion-limited chemical reactions provide test cases of particular interest,
since their evolution {at long times} is dominated by
fluctuations on all time- and length-scales \cite{Matt98,benA00,Henkel08,Kapr10}.
Here, we shall concentrate on the diffusion-limited coagulation process of a single
species of particles, $A$, which can freely diffuse on an underlying lattice and upon encounter undergo a reaction
$A+A\to A$. At the level of a mean-field description via a kinetic equation $\partial_t \rho = - \lambda \rho^2$
for the mean particle-density $\rho=\rho(t)$, one finds $\rho(t)\sim \rho_{\infty} t^{-1}$
at long times \cite{Tous83}. {This algebraic decay is
a clear indication that the system is {\em exactly} at a critical point.}
{The independence of {the amplitude $\rho_{\infty}$ in}
this result from the initial density $\rho(0)$ is an example of  
universality.\footnote{{By `universality', we mean that the long-time behaviour of the system should only depend on a few
`essential' ingredients, which can be specified through the renormalisation group,
and should be independent of the other `details' of the model. Examples of such `details' could
be the coordination number of a regular lattice or the initial particle-density.}}}
However, in spatial dimensions $d\leq 2$, the kinetics of this reaction is {\em anomalous}, since the decay behaviour is different
from the mean-field behaviour and rather becomes $\rho(t)\sim t^{-d/2}$ for $d<2$ and $\rho(t)\sim t^{-1}\ln t$ for $d=2$.
In table~\ref{tab1}, we list experimental examples where 
anomalous kinetics of diffusion-limited reactions has been observed
for effectively one-dimensional systems. Clearly, the $1D$ decay exponent $\alpha$ turns out to be close to the exact,
fluctuation-dominated result $\alpha=\demi$ and is far from the mean-field expectation $\alpha_{\rm MF}=1$.
{The independence, both of the decay exponent $\alpha$ as well as of the scaling amplitude $\rho_{\infty}$,
of the initial density, has also been checked in some of these experiments \cite{Kroon93,Allam13}.}

\begin{table}[b]
\begin{center}\begin{tabular}{|ll|lr|} \hline
~material~      & ~~~$\alpha$  & \multicolumn{2}{c|}{ References } \\ \hline
C$_{10}$H$_8$   & $0.52 -0.59$ & {\scriptsize Prasad and Kopelman (1989)} ~~   & \hfill \cite{Pras89}     \\
P1VN/PMMA       & $0.47(3)$    & {\scriptsize Kopelman {\it et al.} (1990)}    & \hfill \cite{Kopelman90} \\
TMMC            & $0.48(4)$    & {\scriptsize Kroon {\it et al.} (1993)}       & \hfill \cite{Kroon93}    \\
HipCO nanotubes & $0.5$        & {\scriptsize Russo {\it et al.} (2006)}       & \hfill \cite{Russo06}    \\
CoMoCAT         & $0.5$        & {\scriptsize Srivastava and Kono (2009)}~~    & \hfill \cite{Sriv09,Mura09}  \\
HiPco nanotubes & $0.51(3)$    & {\scriptsize Allam {\it et al.} (2013)}       & \hfill \cite{Allam13}    \\ \hline
\end{tabular}\end{center}
\caption[tab1]{Experimentally measured decay exponents $\alpha$ of the mean particle-density $\rho(t)\sim t^{-\alpha}$
from diffusion-limited exiton kinetics in effectively $1D$ systems. \label{tab1}}
\end{table}

On the other hand, in the opposite case of reaction-limited processes, which arises for example for well-stirred systems,
mean-field descriptions are adequate.
{Rigorous upper and lower bounds show that allowing space-dependence of the density, viz. $\rho=\rho(t,\vec{r})$,
does not lead to a long-time behaviour of the spatially averaged density $\rho(t)$
different from that of a well-mixed system  \cite[p. 193]{Henkel08}.}
Therefore, it is of interest to
study the crossover between the diffusion-limited and reaction-limited extreme cases of simple kinetic models, especially as
this crossover has already been studied experimentally \cite{Allam13}.

\begin{figure}[htbp]
\begin{center}
\includegraphics[width=.5\hsize]{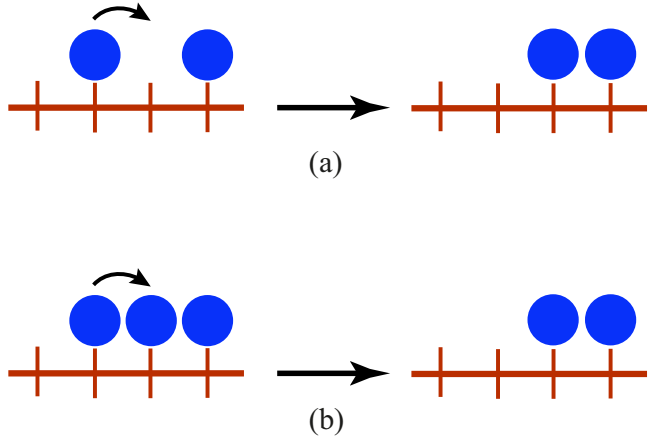}
\end{center}
\caption[fig2]{Movement of single particles of the coagulation-diffusion process:
(a) diffusion movement, during which the total number of particles does not change $A + \emptyset \to \emptyset + A$, and
(b) coagulation movement during which the number of particles {is} reduced  $A +A \to \emptyset + A$.}
\label{fig2}
\end{figure}
Clear and non-ambiguous results of the crossover {from the} diffusion-limited {to the} reaction-limited collective behaviour
are best obtained in the context of exactly solvable models.
{Here, we shall study the {\em coagulation-diffusion process} of particles of a single
species $A$. The model is formulated on a lattice whose sites can be either empty ($\emptyset$) or occupied by a single particle ($A$).
The dynamics of the model is described in terms of uncorrelated random jumps of a single particle.
If upon a jump a particle arrives at an empty site, it is placed there.
However, if it arrives at an occupied site, it is removed from the system with probability one.
This gives two kinds of two-sites microscopic processes,
namely {\em{diffusion}}  $A + \emptyset \to \emptyset + A$ and {\em{coagulation}}  $A +A \to  \emptyset + A$,
as illustrated in figure~\ref{fig2} for particles on a chain.
This kind of nearest-neighbour interactions can be defined on lattices in any spatial dimension.}

\begin{figure}[htbp]
\begin{center}
\includegraphics[width=.5\hsize]{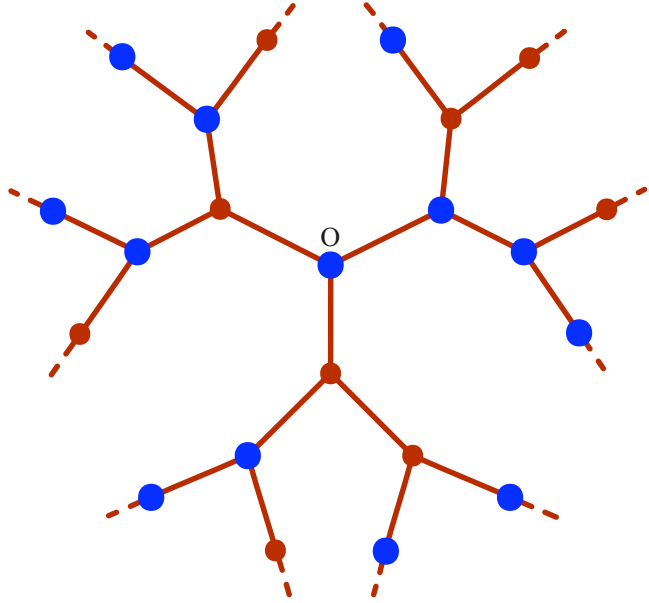}
\end{center}
\caption{Cayley tree with $q=3$ branches and $\ell=3$ generations, starting from the central site O.
The sites of the lattice can be either empty (small brown circles) or occupied by particles of species A (large blue circles).}
\label{fig1}
\end{figure}

{Next, we define the lattice on which we want to study the coagulation-diffusion process.}
In figure~\ref{fig1}, we show the first three generations of a {\em Cayley tree} \cite{Baxter82,Ostilli12}.
A Cayley tree is created from a central site O.
In the first generation, $q$ distinct neighbours are attached to the site $O$.
{Later generations are obtained by induction from the already existing tree at generation $\ell\geq 1$.}
In the generation $\ell+1$, to each of the sites of the generation $\ell$ one attaches $q-1$ distinct neighbours,
such that the generation $\ell$ has $q(q-1)^{\ell-1}$ sites.
After $\ell$ generations, the Cayley tree has ${c_\ell=}q\left[(q-1)^{\ell}-1\right]/(q-2)$ sites \cite{Baxter82}.
{A number of sites on  a distance $\ell$ from a center in a regular lattice is proportional to the volume:
$c_\ell\sim \ell^d$. We can calculate the quantity $\lim_{\ell\to\infty} \ln c_\ell/ \ln \ell=d$
which can be considered as a definition of the dimensionality.
For a hyper-cubic lattice, this limit is finite, while the limit of this ratio on a Cayley tree is easily seen to go to infinity.}
Hence the Cayley tree can be considered to be of infinite dimension \cite{Baxter82}.
The {\em Bethe lattice} \cite{Baxter82,Ostilli12} is {defined} as the interior of the infinite Cayley tree ($\ell\to\infty)$,
after infinitely many generations.
Each site of the Bethe lattice has exactly $q$ nearest neighbours and there is no boundary.
As {a} Cayley tree, the Bethe lattice is also considered being infinite-dimensional \cite{Baxter82,Ostilli12}.
{A well-known theorem \cite{benA06,Rozikov08} states}  that
{\it on the Bethe lattice, a connected cluster of $n$ sites has $n(q-2)+2$ neighbours, independently of the shape of the cluster}.
Both the Cayley tree and the Bethe lattice are widely used in the context
of analytical studies of spin systems and of different chemical reactions
\cite{Turban80,Igloi88,Majumdar93,Abad04,benA06,Matin07,Ali09,Chatelain12,Khorr14,Matin15,Turban17,Dudka17},
random and cooperative sequential adsorption \cite{Cadilhe04,Cadilhe07,Chatelain12,Mazilu12} or branched polymers \cite{Henkel96,Rios01}.

{We want to study the long-time behaviour of the coagulation-diffusion process
on the Bethe lattice and shall compare results with what is known
of the model's behaviour on a chain. Indeed, on a chain the coagulation-diffusion process is diffusion-limited and
can be treated exactly through the well-known method of empty intervals,
introduced by ben-Avraham, Burschka and Doering \cite{benA90,benA00}.
The method has since been substantially generalised
\cite{Racz85,Spouge88,Krebs95,Dahmen95,Rey97,Masser00,Abad00,Henkel01,Yuste01,Yuste02,Abad02,Khorr03},
\cite{Agha05,Muna06a,Muna06b,Durang10,Durang11,Durang14,Fortin14,Fortin17,LeVot18}.
Since the Bethe lattice is infinite-dimensional,
a reaction-limited behaviour should be expected for the same model on the Bethe lattice.
By varying the number of nearest neighbours $q$ continuously from $q=2$ (chain) to $q>2$ (Bethe lattice),
the crossover from the diffusion-limited to the reaction-limited long-time behaviour can be examined.}

\begin{figure}[tb]
\begin{center}
\includegraphics[width=.6\hsize]{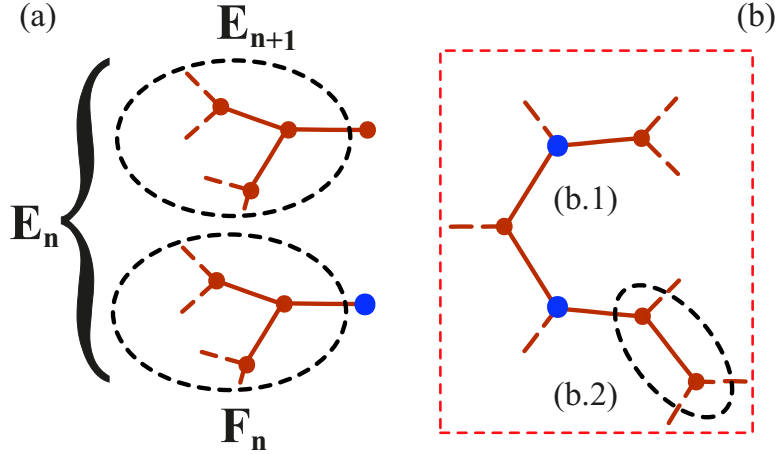}
\caption[fig3]{The empty-interval method on the Bethe lattice, with occupied (blue circles) and empty sites (brown circles) and
with schematically depicted probabilities to find clusters of empty sites ({dashed} ovals).
(a) $E_{n+1}$ is the probability that a connected cluster with $n+1$ sites is empty ({dashed} oval).
$F_{n}$ is the probability that a connected cluster $\mathscr{C}_n$ with $n$ sites is empty and has one occupied nearest neighbour.
If that neighbouring site is connected to $\mathscr{C}_n$ via a single
link, and since $E_{n+1}$ is the probability that the empty cluster $\mathscr{C}_n$ has an empty nearest neighbour,
then $E_n = E_{n+1} + F_n$ and eq.~(\ref{1.2}) holds true.
(b) The cases with the two possible configuration of $3$-clusters (red {dashed} box):
the event (b.1) cannot be presented in the same way as the case (a),
since here empty sites do not form connected cluster as opposed to the event (b.2) \cite{benA06}. \label{fig3}}
\end{center}
\end{figure}

{The method of empty intervals has been generalised to the Bethe lattice, at least approximatively \cite{benA06}.
{For illustration,} figure~\ref{fig1} shows a portion of the Cayley tree with $q=3$ and occupied and empty sites.
Following \cite{benA06}, one considers a connected cluster $\mathscr{C}_n$ with $n$ empty sites
{(schematically shown in figure~\ref{fig3}(a) bounded by dashed lines)} and one defines the
time-dependent probability $E_n(t)$ that the cluster $\mathscr{C}_n$ contains no particles at time $t$.
Since $E_n$ only changes when a particle hops
into the cluster, and recalling that the number of neighbouring sites of $\mathscr{C}_n$ depends only on $n$,
but not on its shape \cite{benA06,Rozikov08}, $E_n(t)$ is completely described by $n$.
In addition, one defines the 
probability $F_n(t)$
that a connected cluster $\mathscr{C}_n$ contains no particles and that one of its neighbouring sites is occupied.}
Then one has the equation of motion, for all $n\geq 1$ \cite{benA06}
\BEQ \label{1.1}
\frac{\D}{\D t} E_n(t) = \frac{n(q-2)+2}{q}\, \digamma \left( F_{n-1}(t) - F_n(t)\right),
\EEQ
with the boundary conditions $E_0(t)=1$ and
$\lim_{n\to\infty}E_n(t)=0$.\footnote{{This boundary condition arises since
the last particle in the system cannot decay \cite{benA00,benA06}.}}
Herein, $\digamma$ is the hopping rate of a particle to a nearest-neighbour site.
By the model's definition, if a particle attempts to hop onto an occupied site, it disappears with probability one.
The system of equations is closed by admitting the {\em ben-Avraham-Glasser approximation} \cite{benA06}
\BEQ \label{1.2}
F_n(t) = E_n(t) - E_{n+1}(t).
\EEQ
Eq.~(\ref{1.2}) is tacitly admitted in \cite{Ali09,Khorr14,Matin15}.
For $q=2$, eqs.~(\ref{1.1},\ref{1.2}) reduce to the exact equations of motion of the empty-interval method on the chain.
{For $q>2$, {(\ref{1.2}) neglects the cases when an occupied site is embedded in a cluster of empty sites, splitting it into
two (or more) pieces. Therefore  in this case }(\ref{1.2}) is an uncontrolled approximation.}
{However, according to ben Avraham and Glasser \cite{benA06},
the method is not only exact for $q=2$, but also for $q\to\infty$.
Furthermore, they argue that even at intermediate values such as $q\approx 3$,
the approximation (\ref{1.2}) should also produce good results.
The validity of eq.~(\ref{1.2}) is further illustrated in figure~\ref{fig3}.}

{As initial state, we consider a random distribution of particles such that a site is occupied with probability $p$.
Then the empty-cluster probability is}
\BEQ \label{1.3}
E_n(0) = \left( 1 - p \right)^n.
\EEQ
The particle density is obtained as $\rho(t) = 1-E_1(t)$.

Explicit analytical insight can be obtained by bringing eq.~(\ref{1.1}) with (\ref{1.2}) to the continuum limit.
{It was checked {in \cite{Abad00,Abad02,Yuste02,Durang10} that the universal long-time behaviour is the same}
on the discrete chain and for
the continuum limit. On the other hand, the non-universal short-time behaviour is different.}
With a lattice constant $a$, set $x=na$ such that $E_n(t) \to E(t,x)$.
To take the continuum limit,
{we let $\digamma=\digamma(a)$ and $q=q(a)$. We then take the limit $a\to 0$, and simultaneously}
$\digamma\to\infty$ and $q-2\to 0$ such that the limits
\BEQ \label{1.4}
\mu := \lim \frac{2}{q}\, \digamma a^2 \;\; , \;\;
\lambda := \lim \frac{q-2}{q}\, \digamma a
\EEQ
exist.\footnote{Herein, $\lambda=0$ reduces to the chain, while the opposite limit $\lambda\to\infty$
gives the behaviour of the Bethe lattice.\\
{A possible way to achieve this is by setting $\digamma(a)=\mu a^{-2}$ and $q(a)=2+\frac{2\lambda}{\mu}a$.}}
Then, in the ben-Avraham-Glasser approximation, from (\ref{1.1},\ref{1.2}) one has on the Bethe lattice the differential equation
\BEQ \label{1.5}
\frac{\partial}{\partial t} E(t,x) = \left( \lambda x + \mu\right) \frac{\partial^2}{\partial x^2} E(t,x) \;\; , \;\;
E(t,0) =1 \;\; , \;\; E(t,\infty) = 0,
\EEQ
where the boundary conditions are already included. The initial condition (\ref{1.3}) becomes $E(0,x) = e^{-cx}$,
where $c$ is the initial concentration of a set of uncorrelated particles \cite{Durang10}.
The sought particle-density is \cite{benA90,benA00}
\BEQ \label{1.6}
\varrho(t) = - \left.\frac{\partial}{\partial x} E(t,x) \right|_{x=0}.
\EEQ
{Therefore, in the continuum limit the coordination number
$q-2 \simeq \frac{2\lambda}{\mu} a\to 0$ becomes infinitesimally close to $q=2$.
Hence, we are infinitesimally close to the chain where (\ref{1.2}) is applicable.
It follows that {\it the ben Avraham-Glasser approximation should become exact in the continuum limit (\ref{1.4}),}
although it is only approximate on the discrete Bethe lattice.
In the continuum limit, we choose the dimensions of time and space such that $\mu>0$ becomes a dimensionless constant.
Then the dimensionful parameter $\lambda$ describes the crossover from the diffusion-limited case (chain) when
$\lambda=0$ to the reaction-limited case (Bethe lattice) when $\lambda\to\infty$.}

{In \cite{benA06}, eq.~(\ref{1.5}) was studied by setting $\mu=0$ {from the outset \cite[eq. (13)]{benA06}},
which in the continuum limit is only possible when $q>2$ is being kept fixed.
Then, a scaling ansatz $E(t,x)=\Phi(x t^{-1/z})$ was tried.
Since $x$ has the dimension of a length, the exponent $z$ can be interpreted as a dynamical critical
exponent.\footnote{{A scaling form $E(t,x)=\Phi(x t^{-1/z})=\Phi(x/L(t))$ always
defines a time-dependent length-scale $L(t)\sim t^{1/z}$ which in turn defines the dynamical exponent $z$
\cite{Matt98,Kapr10,Henkel08}. Computation of correlators on the chain shows that $x$
has the same scaling dimension as a spatial length \cite{Durang10}.}}} The chosen ansatz lead to $z=1$ and the explicit scaling function
$\Phi(u)=\exp\left(-\frac{q}{q-2} u\right)$ \cite{benA06}.
{However, their result appears problematic for several reasons given below.
\begin{enumerate}
\item The singularity for $q\to 2$ prevents a smooth crossover from the Bethe lattice to the chain.
\item A dynamical exponent $z=1$ would imply {\it ballistic transport} across the Bethe lattice,
in clear contrast to the diffusive motion of the single particles. Such a result, if indeed true, would be extremely surprising
{and be in contradiction to the derivation of the equation of motion (\ref{1.1}) from the random
hopping of single particles between nearest-neighbour sites on the lattice}.
\item Observables such as the density are found by calculating derivatives of $E(t,x)$ at $x=0$, see (\ref{1.6}).
This means that in the expression $\lambda x+\mu$ which arises in (\ref{1.5}), {the term $\mu$ {\em cannot} be considered
negligible with respect to $\lambda x$}.
\end{enumerate}
We conclude: simply setting $\mu=0$ in (\ref{1.5}) is not legitimate and a full analysis with $\mu>0$ must be carried out.
Indeed, we shall show in section~3 that keeping $\mu>0$ is important and leads to a dynamical exponent $z=2$,
consistent with diffusive motion of the single particles.
Also, the crossover from the chain to the Bethe lattice can be derived explicitly.}

{However, no simple scaling ansatz to solve eq.~(\ref{1.5}) could be found.
In order to understand how this might arise, an useful intermediate problem is to consider the stationary state
of the coagulation-diffusion process in the presence of a {\em stochastic reset}. This concept was introduced by Evans and
Majumdar \cite{Evans11a,Evans11b} for the example of brownian motion of a single particle.
The dynamics proceeds in small time intervals $\Delta t$. At each time step,
either the particle is reset to the origin with probability $r\Delta t$
or else it makes a step of usual brownian motion, with probability $1-r\Delta t$. Herein,
the parameter $r>0$ is called the {\em reset rate}.
The associated master equation is a modified form of a diffusion equation and leads
to a non-gaussian stationary state \cite{Evans11a,Evans11b,Evans14}.
{}From the analysis of search algorithms,
it can be shown that a stochastic reset may accelerate the relaxation of the statistical system
to a new kind of non-equilibrium stationary state \cite{Evans11a,Evans11b,Evans14,Fuchs16,Falcao17,Montero17,Roldan17},
with applications to RNA polymerase \cite{Roldan16}.}

{These concepts can be extended to many-body problems,
such as the $1D$ coa\-gu\-la\-tion-diffu\-sion process described by the empty-interval method \cite{Durang14}.
Here, we characterise the reset by a prescribed distribution $S(x)$ of empty intervals of
size $x$.\footnote{For example, one might consider $S(x)=E(0,x)$ which would
describe a reset to the initial state. {$S(x)$ being an empty-cluster probability, it naturally obeys $S(0)=1$ and $S(\infty)=0$.}}
The particle-density of the resetting state is $c :=-\left.\partial_x S(x)\right|_{x=0}$.
In analogy with a reset in brownian
motion described above, one either resets at each time step the entire system to the state described by $S(x)$,
with probability $r\Delta t$,
or else performs a standard step of the coagulation-diffusion process, with probability $1-r\Delta t$.
Taking a continuum limit greatly simplifies the analysis.
Here, we generalise our earlier treatment of the $1D$ coagulation-diffusion process \cite{Durang14} to the Bethe lattice.
We focus on the continuum limit and on the stationary state,
with the stationary empty-cluster probability $E(x) := \lim_{t\to\infty} E(t,x)$.
Generalising the stationary limit of (\ref{1.5}), we have}
\BEQ \label{1.7}
\left( \lambda x + \mu\right) \frac{\partial^2}{\partial x^2} E(x) - r E(x) + r S(x) = 0 \;\; , \;\;
E(0) =1 \;\; , \;\; E(\infty) = 0.
\EEQ
Herein, the reset rate $r$ takes the role of a control-parameter describing the distance from the free coagulation-diffusion process,
since the relaxation time towards the stationary state
diverges as $r\to 0$
\cite{Evans11a,Evans11b,Evans14,Durang14}.\footnote{{We assume throughout that $r$ is small enough that the continuum limit is
applicable.}} Clearly, the stationary density $\varrho=-\left.\partial_x E(x)\right|_{x=0}$.
{By dimensional analysis, the reset rate defines a time-scale $t_{\rm r}\sim 1/r$,
and an associated length scale $\ell_r \sim t_{\rm r}^z$} such that on distances $\ell \ll \ell_r$ the correlations coming from the
usual dynamics without a reset are found while for distances $\ell \gg \ell_r$
the correlations of the resetting state are maintained \cite{Durang14}.
The parameter $\lambda$ controls as before the nature of the dynamics without the
reset and we shall appeal to conventional crossover scaling theory \cite{Lueb04,Henkel08} in $\lambda$
for the interpretation of the results. Herein, the reset will serve as a guide for the interpretation of the non-steady-state dynamics.

This work is organised as follows. In section~2, we give the solution of the stationary state problem (\ref{1.7}) with a reset.
{On the Bethe lattice, a logarithmic modification of the scaling behaviour with respect to the expectation from the chain will be derived.}
In section~3, we study the time-dependent equation (\ref{1.5}), {without a reset,} which in Laplace space is analogous to (\ref{1.7}).
{The logarithmic modifications of the scaling behaviour in Laplace space (as obtained before from the stationary state with a reset)
requires a careful mathematical analysis in order to invert the Laplace transform correctly.}
Section~4 presents the crossover scaling and we conclude in section~5.

\section{Stationary-state behaviour with a reset}

Before presenting the analysis of  the exact solution of the coagulation-diffusion model with a stochastic reset,
let us briefly discuss the mean-field result. The dimensional analysis leads to the following.
If $\Lambda$ denotes a length scale of reference, then from the definitions and eq.~(\ref{1.7}) we have the scaling dimensions
$[\lambda]=\Lambda^{-1}$, $[\mu]=\Lambda^0=1$, $[c]=\Lambda^{-1}$ and $[r]=\Lambda^{-2}$.
Clearly, the particle-density $\varrho$ should have the same dimension as the concentration $c$.
Any other dependence can only enter through functions of dimensionless arguments,
hence, obviously, a mean-field particle-density must be of the form
\BEQ \label{2.1CM}
\varrho_{\rm MF} = c f_{\rm MF}\left( \frac{r}{\lambda c}, \frac{\lambda}{c} \right).
\EEQ
Furthermore, for small reset rates $r\to 0$, the stationary density should
be independent of the density $c$ of the state to which the reset is done.
{Hence, for $r$ small enough, the scaling function $f_{\rm MF}$ should become independent of its second argument.}
This fixes the form of the mean-field scaling function $f_{\rm MF}$ for a small first argument, hence $\varrho_{\rm MF}\sim r/\lambda$.

We now turn to the exact solution of the coagulation-diffusion model with a stochastic reset.
The equation of motion is (\ref{1.7}), with
the relevant boundary conditions, where $S(x)$ characterises the empty intervals of the reset configuration.
It obeys the boundary conditions $S(0)=1$ and $S(\infty)=0$ \cite{Durang14}. {The specific form of $S(x)$ will be given below.}

Following \cite{Durang14,Kamke77}, a  basis of solutions of the associated homogeneous equation is spanned by the functions
$\lambda^{-1}\sqrt{r(\lambda x+ \mu)\,}\,I_1\left(\frac{2\sqrt{r}}{\lambda}\sqrt{\lambda x+ \mu}\,\right)$ and
$\lambda^{-1}\sqrt{r(\lambda x+ \mu)\,}\,K_1\left(\frac{2\sqrt{r}}{\lambda}\sqrt{\lambda x+ \mu}\,\right)$,
where $I_1, K_1$ are the modified Bessel functions of order $1$ \cite{Abra65}.
By the method of variation of the constants \cite{Kamke77}, we find
\bb
\lefteqn{E(x) = A \xi(x)\,K_1\left(2 \xi(x)\right) +
B \xi(x)\,I_1\left(2 \xi(x)\right)} \\ \nn
&\!\!\!\!+& \!\!\!\!2\sqrt{r}\,\xi(x)\,K_1\left(2\xi(x)\right)
    \int_0^x \frac{\sqrt{r}\, I_1\left(2\xi(z)\right)}{\lambda\xi(z)} S(z) \D z
    + 2\sqrt{r}\,\xi(x)\,I_1\left(2\xi(x)\right)
    \int_{x}^{\infty} \frac{\sqrt{r} K_1\left(2\xi(z)\right)}{\lambda\xi(z)} S(z) \D z,
\ee
where we have used the short-hand notation $\xi(x) := \sqrt{r(\lambda x + \mu)\,}\,/\lambda$.
Taking the boundary conditions into account, we see that $B=0$ and
\bb
A = \frac{1}{u\,K_1\left(2 u\right)}
- 2\frac{r}{\lambda}\,\frac{I_1(2u)}{K_1(2u)}
\int_0^\infty \frac{K_1\left(2\xi(z)\right)}{\xi(z)} S(z) \D z,
\ee
with $u=\sqrt{r\mu}/\lambda$.
Hence, the general solution of (\ref{1.7}), still for $S(x)$ arbitrary, is given by
\begin{eqnarray}
\label{2.3}
E(x) &=& \frac{\xi(x)}{u}
         \frac{K_{1}\left(2 \xi(x)\right)}{K_{1}\left(2 u\right)}  \nonumber \\
\nonumber \\
& & -{2\sqrt{r}}{\xi(x)}
    \frac{I_{1}\left(2 u\right)}{K_{1}\left(2 u\right)}
    K_{1}\left(2\xi(x)\right)
    \int_{0}^{\infty} \!\D z\: \frac{\sqrt{r}K_{1}\left(2 \xi(z) \right)S(z)}{\lambda \xi(z)}
\nonumber \\
& & +{2\sqrt{r}}{\xi(x)} I_{1}\left(\frac{2\xi(x)}{\sqrt{r}}\right)
    \int_{x}^{\infty} \!\D z\: \frac{\sqrt{r}K_{1}\left(2 \xi(z) \right)S(z)}{\lambda \xi(z)}
\nonumber \\
& & +{2\sqrt{r}}{\xi(x)} K_{1}\left(\frac{2\xi(x)}{\sqrt{r}}\right)
    \int_{0}^{x} \!\D z\: \frac{\sqrt{r}I_{1}\left(2 \xi(z) \right)S(z)}{\lambda \xi(z)}.
\end{eqnarray}
The stationary particle-density $\varrho$ is
\begin{eqnarray}
\varrho     &=& - \left.\frac{\partial E(x)}{\partial x}\right|_{x=0}
             =  \sqrt{\frac{r}{\mu}}\, \frac{K_{0}\left(2 u\right)}{K_{1}\left(2u\right)}
                - \frac{r}{\sqrt{\mu}} \frac{1}{K_{1}\left(2u\right)}
                \int_{0}^{\infty} \!\D z\: \frac{\sqrt{r}\,K_{1}\left(2 \xi(z) \right)S(z)}{\lambda \xi(z)}.
\label{2.4}
\end{eqnarray}
Using 
the identity \cite[eq. (9.6.27)]{Abra65}, we can express $K_1$ as follows
\BD
\frac{\sqrt{r}K_{1}[2\xi(x)]}{\lambda\xi(x)} = - \frac{1}{\sqrt{r}} \frac{\partial K_{0}[2\xi(x)]}{\partial x},
\ED
and integrating eq.~(\ref{2.4}) by parts, we obtain the more simple form
\begin{eqnarray}
\label{2.5}
\varrho = \sqrt{\frac{r}{\mu}} \frac{1}{K_{1}\left(2 u\right)}
          \int_{0}^{\infty} \!\D z\: K_{0}\left(2\xi(z)\right)\left( - \frac{\D S(z)}{\D z}\right).
\end{eqnarray}

For explicit examples, we shall consider here a reset to a configuration of uncorrelated particles, with concentration $c$.
Then $S(x) = e^{- c x}$. {From now on, we shall restrict to this special case.}

In particular, in the limit $\lambda\to 0$ we obtain the results for the chain,
indeed reproduce the known result \cite{Durang14} and cast it in a scaling form
\BEQ \label{2.6}
\left.\varrho\right|_{\lambda \rightarrow 0} = \frac{c \sqrt{r/\mu}}{c + \sqrt{r/\mu}}
= c \frac{\sqrt{r/(\mu c^2)\,}}{1+\sqrt{r/(\mu c^2)\,}} = c\, \mathscr{Q}(v).
\EEQ
with the scaling variable $v=\sqrt{r/(\mu c^2)\,}$ and the scaling function $\mathscr{Q}(v)=v/(1+v)$.
Similarly, for arbitrary $\lambda>0$ eq.~(\ref{2.5}) can be cast into a scaling form
\BEQ \label{2.7}
\varrho = c \mathscr{P}\left( \frac{\sqrt{r\mu\,}}{\lambda} , \frac{c\mu}{\lambda} \right) \;\; , \;\;
\mathscr{P}(u,w) = \frac{u}{K_{1}\left(2u\right)}\int_{0}^{\infty} \!\D y\: K_{0}\left(2 u \sqrt{y+1\,}\,\right) e^{-w y},
\EEQ
with the scaling variables $u = \sqrt{r \mu}/\lambda$ and $w=c\mu/\lambda$ (such that $v=u/w$).
Asymptotically, we have, where $C_E\simeq 0.5772\ldots$ is Euler's constant
\BEQ \label{2.8}
\mathscr{P}\left(u,\frac{u}{v}\right) \simeq \left\{
\begin{array}{ll} \frac{v}{v+1}                  & \mbox{\rm ~~;~ if $u\to\infty$}, \\[0.12truecm]
                  -uv\left( \ln(uv) + C_E\right) & \mbox{\rm ~~;~ if $u\to 0$}.
\end{array} \right.
\EEQ
{The limit case $u\to\infty$ (or $\lambda\to 0$) reproduces the known result (\ref{2.6}) of the chain.
The other limit $u\to 0$ (or $\lambda\to\infty$), however, gives an unexpected behaviour of the model on the Bethe lattice.}

Eq.~(\ref{2.8}) is derived as follows: first for $u\to\infty$, one uses the asymptotics
$K_{\nu}(z)\simeq \sqrt{\pi/2z\,}\, e^{-z} \left(1+\mbox{\rm O}(1/z)\right)$ such that
\BD
\mathscr{P}\left(u,\frac{u}{v}\right)
           \simeq u e^{2u} \int_0^{\infty} \!\D y\: \frac{\exp\left( -2u\sqrt{y+1\,}\,-uy/v\right)}{(y+1)^{1/4}}
           \simeq u\int_0^{\infty} \!\D y\: e^{-u(1+1/v)y} = \frac{v}{v+1},
\ED
where we used the fact that the main contribution to the integral comes from values $y\ll 1$. Second, for $u\to 0$, we rewrite the integral as follows
\BEA
\lefteqn{\mathscr{P}\left(u,\frac{u}{v}\right) = \frac{u e^{u/v}}{K_1(2u)}
\left[ \int_0^{\infty} \!\D z\: K_0\left(2u z^{1/2}\right)e^{-uz/v} - \int_0^{1} \!\D z\: K_0\left(2u z^{1/2}\right)e^{-uz/v} \right]}
\nonumber \\
&\simeq& \frac{u e^{u/v}}{K_1(2u)}
\left[ \frac{v}{2u} e^{uv} \Gamma(0,uv) +\int_0^1 \!\D z\: \left[ \ln\left( u z^{1/2}\right) + C_E \right] e^{-uz/v} \right]
\nonumber \\
&=& \frac{u e^{u/v}}{K_1(2u)} \left[ \frac{v}{2u} e^{uv} \Gamma(0,uv) + \left( \ln u + C_E\right)\frac{\left( 1 - e^{-u/v}\right)}{u/v} +
\demi\left( C_E + \ln \frac{u}{v} + \Gamma\left(0,\frac{u}{v}\right)\right) \right]
\nonumber \\
&\simeq& uv \left( -C_E - \ln(uv)\right) + \mbox{\rm O}(u^2),
\nonumber
\EEA
where in the second line, the first integral is evaluated with \cite[(3.16.2.2)]{Prud4},
where $\Gamma(0,x)$ is an incomplete Gamma function
\cite{Abra65},\footnote{$\Gamma(0,x)=\mbox{\rm E}_1(x)=\mbox{\rm Ei}(1,x)=-\mbox{\rm Ei}(-x)$
can be expressed as an exponential integral, {in different notations, see e.g.} \cite{Abra65,Prud1} {and the Maple handbook.}}
and in the second term, the leading contribution to the Bessel function $K_0$ for small arguments was estimated
with \cite[(9.6.13)]{Abra65}. The last integral was evaluated in the third line with \cite[(1.6.10.2)]{Prud1},
followed by an expansion in $u$ to leading order, using \cite[(6.5.15,5.1.11)]{Abra65}.
\hfill q.e.d.

Finally, going back to the original variables, we find from (\ref{2.8}) the following scaling behaviour for small reset rates $r\ll 1$
\BEQ \label{2.10}
\varrho \simeq \left\{ \begin{array}{lll}
\sqrt{r\,}/\mu                                                 & \mbox{\rm ~~;~ if $\lambda\to 0$}     & \mbox{\rm chain}, \\[0.12truecm]
\frac{r}{\lambda} \left[ \ln \frac{\lambda c}{r} - C_E \right] & \mbox{\rm ~~;~ if $\lambda\to\infty$} & \mbox{\rm Bethe lattice}.
                       \end{array} \right.
\EEQ
Indeed, this is of the generic form argued for above in eq.~(\ref{2.1CM})
and we also see that for $r\ll 1$ the resetting concentration $c$ cancels out.
However, we observe on the Bethe lattice a logarithmic correction to the scaling with the reset rate $r$.
This can be interpreted as a correlation effect which distinguishes the
behaviour on the Bethe lattice from a {simple} mean-field treatment.

{In this section, we have studied the properties of the non-equilibrium stationary state which arises from a non-vanishing reset rate. In the next section we consider the time-dependent behaviour of the coagulation-diffusion process on Bethe lattice in the absence of  a reset. We shall see that the equations obtained in this case are analogous to those considered in this section.}

\section{Time-dependent behaviour}

{We now turn to the time-dependent behaviour of the coagulation-diffusion process,
but without a reset.\footnote{{If a reset with fixed rate $r>0$
would also be present, it is easily seen that in Laplace space, the variable conjugate to time is $s+r$,
to be evaluated in the long-time limit $s\to 0$. This would give an exponentially rapid and non-universal relaxation
to the stationary state with a reset, studied above, and would provide no information
on the universal and algebraic long-time relaxation behaviour of the coagulation-diffusion process with $r=0$.}}}
The time-dependent eq.~(\ref{1.5}) is solved via a Laplace transformation.
Writing $\,\lap{f}(s) = \mathscr{L}\left( f(t)\right)(s) = \int_0^{\infty}\!\D t\: e^{-st} f(t)$, the transformed equation of motion
\BEQ \label{3.1}
s\lap{E}(s,x) -E(0,x)=(\lambda x + \mu)\partial_x^2 \lap{E}(s,x)
\EEQ
is almost identical to the stationary equation of motion (\ref{1.7}) for the reset.
The solution therefore proceeds along almost identical lines.
Starting from the basis of the solution of the homogeneous equation, namely
$\sqrt{\lambda x+ \mu}\,K_1\left(\frac{2\sqrt{s}}{\lambda}\sqrt{\lambda x+ \mu}\right)$ and
$\sqrt{\lambda x+ \mu}\,I_1\left(\frac{2\sqrt{s}}{\lambda}\sqrt{\lambda x+ \mu}\right)$
the general solution of the differential equations now reads
\bb
\lap{E}(s,x) &=& A \frac{\lambda \xi(x)}{\sqrt{s}}\,K_1\left(2 \xi(x)\right) +
B\frac{\lambda \xi(x)}{\sqrt{s}}\,I_1\left(2 \xi(x)\right) \\ \nn
& & + \frac{2 \xi(x)}{\sqrt{s}}\,K_1\left(2\xi(x)\right)
    \int_0^x \frac{\sqrt{s}I_1\left(2\xi(z)\right)}{\lambda\xi(z)} E_0(z) \D z \\ \nn
& & + \frac{2 \xi(x)}{\sqrt{s}}\,I_1\left(2\xi(x)\right)
    \int_x^\infty  \frac{\sqrt{s}K_1\left(2\xi(z)\right)}{\lambda\xi(z)} E_0(z) \D z,
\ee
where $\xi(x):=\sqrt{s(\lambda x + \mu)\,}\,/\lambda$ now and $E_0(x)=E(0,x)$ is the initial condition.
The only difference {compared to} the reset from section~2 comes from the boundary conditions which now are
\bb \label{3.3}
\lap{E}(s,0)=1/s \quad {\rm{and}} \quad \lim_{x\rightarrow \infty} \lap{E}(s,x) = 0.
\ee
Hence, $B=0$ and
\bb
A = \frac{1}{s\sqrt{\mu}\,K_1\left(\frac{2\sqrt{s\mu}}{\lambda}\right)}
- \frac{2I_1(\frac{2\sqrt{s\mu}}{\lambda})}{\lambda K_1(\frac{2\sqrt{s\mu}}{\lambda})}
\int_0^\infty \frac{\sqrt{s}K_1\left(2\xi(z)\right)}{\lambda\xi(z)} E_0(z) \D z.
\ee
Using the scaling variables $u = \sqrt{s\mu}/\lambda$ and $v = x \lambda/\mu$, that is $\xi(x)=u\sqrt{1+v}$,
we obtain the scaled empty-interval probability, in the following form
\bb
\lap{E}(u,v) &=& \frac{\sqrt{1+v}}{s} \left[ \frac{K_1\left(2u\sqrt{1+v}\right)}{K_1\left(2u \right)} \right.\\ \nn
& & \quad- 2u^2 K_1\left(2u\sqrt{1+v}\right)\frac{I_1\left(2u \right)}{K_1\left(2u \right)}
    \int_0^\infty \frac{K_1\left(2u\sqrt{1+z}\right)}{\sqrt{1+z}} E_0\left(\frac{z\mu}{\lambda}\right) \D z \\ \nn
& & \quad+ 2u^2K_1\left(2u\sqrt{1+v}\right) \int_0^{\lambda x/\mu} \frac{I_1\left(2u\sqrt{1+z}\right)}{\sqrt{1+z}}
    E_0\left(\frac{z\mu}{\lambda}\right) \D z \\ \nn
& & \quad\left.+ 2u^2I_1\left(2u\sqrt{1+v}\right)
    \int_{\lambda x/\mu}^\infty \frac{K_1\left(2u\sqrt{1+z}\right)}{\sqrt{1+z}} E_0\left(\frac{z\mu}{\lambda}\right) \D z \right].
\ee

The time-space scaling is described by the scaling variable $u v = \sqrt{\mu s\,}\, x$.
{This scaling variable is independent} of the parameter $\lambda$ which describes the crossover
between the chain and the Bethe lattice. Since $\mu>0$ is a fixed dimensionless constant,
we read off the dynamical exponent $z=2$. This means that on the chain as well as on the Bethe lattice the
transport is diffusive.

\subsection{Density}

The Laplace-transformed particle-density is $\lap{\varrho}(s)=-\left.\partial_x \lap{E}(s,x)\right|_{x=0}$.
It assumes a scaling form $\lap{\varrho}(s) = s^{-1} f_{\overline{\varrho}}(u)$
where the scaling variable $u=\sqrt{s\mu}/\lambda$. Explicitly
\bb \label{3.6}
\lap{\varrho}(s) = \frac{1}{s} \frac{u}{K_1(2u)} \int_0^\infty \!\D y\: K_0\left(2u\sqrt{y+1\,}\,\right)
                   {\cal P}_0 \left(\frac{y\mu}{\lambda}\right)
\ee
with ${\cal P}_0(x)=-\partial_x E_0(x)$ is the initial probability of a connected empty interval of size $x$,
with a nearest neighbour occupied \cite{Durang10}.
We choose as initial condition a set of uncorrelated particles of concentration $c$, hence ${\cal P}_0(x) = c \e^{-cx}$.

{}From this last expression, one can extract the limiting behaviour of the scaling function
$f_{\overline{\varrho}}(u)$ for the two cases of interest:
(a) $u \gg 1$ which means that the topology is getting close to the chain ($\lambda \rightarrow 0$),
(b) $u\ll 1$ which means that we should be in the mean-field approximation ($\lambda \rightarrow \infty$).
First, for $u\gg 1$, the asymptotic behaviour of the Bessel function $K_\nu(z) \sim \e^z/\sqrt{z}$ gives
\bb \label{3.7}
\lap{\varrho}(s) \simeq \frac{cu\e^{2u}}{s}\int_0^\infty \!\D y\: \frac{\e^{-2u\sqrt{y+1}-\mathfrak{a}y}}{(y+1)^{1/4}} \;\; , \;\;
\mbox{\rm with $\mathfrak{a}=c\mu/\lambda$}.
\ee
Within this limit, only the region $y\ll 1$ contributes to the integral and we can rewrite it as
\bb
\lap{\varrho}(s) \simeq \frac{cu}{s}\int_0^\infty \!\D y\: \frac{\e^{-uy-\mathfrak{a}y}}{y/4+1}
                 = \frac{cu}{s(\mathfrak{a}+u)} = \frac{\sqrt{s\,}}{c\sqrt{\mu\,}+\sqrt{s\,}}.
\ee
Inverting the Laplace transform, we recover the well-known expression for the particle-density on a chain, {e.g. \cite{Dura10}}
\bb
\varrho(t) = c\,\e^{\mu c^2 t}\erfc\left(c\sqrt{\mu t\,}\,\right)
\ee
with the expected asymptotic behaviour $\varrho(t) \stackrel{t \gg 1}{\rightarrow} \left(\pi \mu t\right)^{-1/2}$.
Second, for $u\ll 1$, we start again from (\ref{3.6}), separate the integration domain and use the short-hand from (\ref{3.7}).
We then shift the integration domain, and in the second term, we also expand the Bessel function for small arguments.
This gives, to leading order in $u$
\BEA
\lap{\varrho}(s) &=& \frac{1}{s} \frac{cu}{K_1(2u)} \left[
\int_{-1}^\infty \!\D y\: K_0\left(2u\sqrt{y+1\,}\,\right) \e^{-ay}
- \int_{-1}^0 \!\D y\: K_0\left(2u\sqrt{y+1\,}\,\right) \e^{-ay}\right]
\nonumber \\
&\simeq& \frac{cu\e^a}{sK_1(2u)}\left[\int_0^\infty \!\D y\: K_0\left(2u\sqrt{y\,}\,\right) \e^{-ay}
+ \int_0^1 \!\D y\: \left(\ln\left(u\sqrt{y\,}\,\right) +C_E\right)\e^{-ay} \right].
\label{3.10}
\EEA
The first integral is known \cite[eq.(3.16.2.2)]{Prud4}
and the second one can be read off from \cite[(1.6.10.2)]{Prud1},\cite[(6.5.15)]{Abra65}
\bb \label{3.11}
\lap{\varrho}(s) \simeq \frac{cu^2\e^a}{sa}\left[\e^{u^2/a}\Gamma(0,u^2/a) +2\ln(u)\left(1-\e^{-a}\right)
+\left(C_E\left(1-2 \e^{-a}\right) -\ln(a)-\Gamma(0,a)\right)\right]
\ee
with the incomplete Gamma-function $\Gamma(0,x)$ \cite{Abra65}. 
When $a\to 0$, we conclude that only the first term in (\ref{3.10}) contributes to the leading order.
Therefore the particle-density should be recast in the original variables as
\begin{subequations} \label{3.12}
\begin{align} \label{3.12a}
\lap{\varrho}(s) &\simeq \frac{\e^{c\mu/\lambda}}{\lambda}\, \e^{{s}/{(\lambda c)}}\:\Gamma\left(0,\frac{s}{\lambda c}\right).
\end{align}
Computing the inverse Laplace transform of (\ref{3.12a}), one recovers the time-dependent particle-density
\begin{align} \label{3.12b}
\varrho(t) &= \frac{c\,\e^{c\mu/\lambda}}{1+t\lambda c} \sim t^{-1},
\end{align}
\end{subequations}
and which would reproduce the asymptotic mean-field behaviour \cite{benA06,Ali09,Khorr14,Matin15}.
Indeed, the mean-field scaling discussed for the reset in the previous section~2 can be
taken over almost unchanged, where now $s$ plays the role previously taken by the reset rate $r$. The only change comes from the
initial condition (\ref{3.3}), such that now
$\lap{\varrho}_{\rm MF}(s) = \frac{c}{s} f_{\rm MF}\left( \frac{s}{\lambda c}\right) \sim \lambda^{-1}$ for small $s$,
which would be equivalent to $\varrho_{\rm MF}(t) \sim t^{-1}$.

However, the derivation of eqs.~(\ref{3.12}) does not properly take into account the presence of logarithmic terms
for small values of $s$ in $\lap{\varrho}(s)$, in the limit of large values of $\lambda$. Indeed, restituting the original variables into
(\ref{3.11}) and then expanding for $\lambda\to\infty$, we have
\BEA
\lap{\varrho}(s) &=& \frac{\e^{c\mu/\lambda}}{2 \lambda}
\left[ \e^{s/(\lambda c)}\,\Gamma\left(0,\frac{s}{\lambda c}\right) \right. \nonumber \\
& & \left. +\ln\left(\frac{\mu s}{\lambda^2}\right)
\:\left(1-\e^{-c\mu/\lambda}\right) +\left( C_E(1-2\e^{-c \mu/\lambda})
       -\ln\frac{c\mu}{\lambda} -\Gamma\left(0, \frac{c\mu}{\lambda}\right)\right)\right]
\nonumber \\
&\simeq& - \frac{\ln s}{\lambda} + \frac{1}{\lambda}\left( \ln(\lambda c)- 2 C_E\right)  + \mbox{\rm O}(\lambda^{-2}).
\label{3.13}
\EEA
This leading logarithmic term $\sim \ln s$ is the analogue of (\ref{2.10}) for the reset.

\subsection{Integrated density}

{An inverse Laplace transformation for the asymptotic form (\ref{3.13}) does not exist. To make progress,
we need the following mathematical result} \cite[ch. XIII.5]{Feller71}.

\noindent
{\bf Definition.} {\it A function $L(t)$ is said to be {\em slowly varying at infinity}, if $L(tx)/L(t)\to 1$ for $t\to\infty$ and for any fixed positive $x$.}

\noindent
Clearly, $L(t)=\ln t$ is an example of a function varying slowly at infinity.

\noindent
{\bf Theorem.} (Hardy-Littlewood-Karamata-Feller) {\it If $F(t)=\int_0^{t}\!\D t'\, f(t')$, $L(t)$ is a function  varying slowly at infinity
and $0\leq\kappa<\infty$, then the following statements are equivalent}
\BEQ \label{3.14}
\lap{f}(s) \stackrel{s\to 0}{\sim} s^{-\kappa} L\left(\frac{1}{s}\right) \mbox{\it ~~ and ~~}
F(t) \stackrel{t\to\infty}{\sim} \frac{1}{\Gamma(\kappa+1)}\, t^{\kappa} L(t).
\EEQ

\noindent \noindent
{\noindent  \underline{Remark}: If $\kappa>0$, and if there is a finite $t_0$ such that $f(t)=\frac{\D F(t)}{\D t}$ exists
and furthermore is monotone for $t_0 < t<\infty$, then asymptotically $f(t) \sim \kappa F(t)/t$, as $t\to\infty$ \cite{Feller71}.
This directly relates the asymptotics of $\lap{f}(s)$ for $s \to 0$ to the one of $f(t)$ for $t\to\infty$, but cannot be extended to $\kappa=0$.}

This suggests to study the integrated density\footnote{Denoted by the capital Greek letter $P$ (rho).} $P(t) := \int_0^{t} \!\D t'\, \varrho(t')$,
instead of $\varrho(t)$. In Laplace space $\lap{P}(s) = s^{-1} \lap{\varrho}(s)$. Therefore, from the above we have
\BEA
\lap{P}(s) &=& \frac{c \sqrt{s\mu}/\lambda}{s^{2} } \frac{1}{K_1\left( 2 \sqrt{s\mu}/ \lambda\right)}
\int_0^{\infty} \!\D z\: K_0\left( \frac{2}{\lambda} \sqrt{ s(\lambda z+\mu)\,}\,\right) \e^{-cz}
\label{3.15}
\\
&=& \frac{c \e^{c\mu/\lambda}\sqrt{s\mu}/\lambda}{ s^{2} K_1\left( 2 \sqrt{s\mu}/ \lambda\right)}
\left[ \frac{\lambda}{2c\mu}\, \e^{s/(\lambda c)}\, \Gamma\left( 0, \frac{s}{\lambda c}\right)
- \int_0^{1} \!\D y\: K_0\left( \frac{2}{\lambda} \sqrt{ \mu s y\,}\,\right) \e^{-c\mu y/\lambda} \right], ~~
\EEA
where \cite[(3.16.2.2)]{Prud4} was used.

While for $\lambda\to 0$ a new discussion  is not necessary, we consider the asymptotic case $\lambda\to\infty$, where from (\ref{3.13}) we have
\BEQ
\lap{P}(s) \simeq -\frac{1}{\lambda} \frac{\ln s}{s} + \frac{\ln(\lambda c) -2 C_E}{\lambda} \frac{1}{s} + \mbox{\rm O}(\lambda^{-2})
\EEQ
and by using \cite[(2.5.1.5),(2.1.1.1)]{Prud5} we have for large times
\BEQ
P(t) \simeq \frac{1}{\lambda}\left( C_E +\ln t \right) + \frac{\ln(\lambda c) -2 C_E}{\lambda} + \mbox{\rm O}(\lambda^{-2})
= \frac{\ln \lambda c t}{\lambda} - \frac{C_E}{\lambda} + \mbox{\rm O}(\lambda^{-2}).
\EEQ
This logarithmic behaviour reproduces the expected mean-field behaviour for the density
\BEQ
\varrho(t) = \frac{\partial P(t)}{\partial t} \sim \frac{1}{\lambda t}
\EEQ
without a logarithmic term in the time-dependence of the density.

\section{Crossover scaling}

In order to describe the crossover behaviour between the coagulation process on the Bethe lattice and on the chain,
we appeal to the corresponding scaling theory \cite{Lueb04,Henkel08}. In the stationary state
with a reset rate $r$, the crossover is described by the change of the parameter $\lambda$ from the chain ($\lambda=0$)
to the Bethe lattice ($\lambda\to\infty$). {In addition, we must take into account the scaling of the reset concentration $c$.}
Then the stationary density should obey the scaling form
\BEQ
{\varrho(r,c,\lambda) = b^{-\beta}\mathscr{R}\left( b r, b^{1/\phi'} c (br)^{-1/\phi}, b^{1/\phi} \lambda\right)
= \lambda^{\beta\phi} \mathscr{R}\left( r \lambda^{-\phi}, c r^{-1/\phi} \lambda^{1-\phi/\phi'}, 1 \right)},
\EEQ
where $b>0$ is a rescaling factor, {$\phi>0$ and $\phi'>0$ are the crossover exponents} and $\mathscr{R}$ is a scaling function.
Writing this, we treat $\mu>0$ as a fixed, dimensionless constant.

{{}From the discussion of the scaling of the stationary state in section~2, we have that $\lambda$ and $c$ should have the same scaling dimensions, which means that $\phi=\phi'$. We then define a dimensionless concentration $\sigma$
of the reset via
\BEQ
c = \sigma r^{1/\phi}
\EEQ
and have the scaling form
\BEQ
\varrho(r,c,\lambda) = \lambda^{\beta \phi} \mathscr{R}\left(r \lambda^{-\phi}, \sigma, 1\right).
\EEQ
}
The crossover between the two scaling regimes, {for all values of the constant $\sigma$}, is described by
\BEQ
{\mathscr{R}(x,\sigma,1)} \sim \left\{
\begin{array}{lll} x^{\beta}          & \mbox{\rm ~~;~ for $x\to \infty$} & \mbox{\rm as $\lambda\to 0$ ~~chain} \\
                   x^{\beta_{\rm MF}} & \mbox{\rm ~~;~ for $x\to 0$}      & \mbox{\rm as $\lambda\to\infty$ Bethe lattice}
\end{array} \right.
\EEQ
and can be traced through the values of the effective exponent {$\beta_{\rm eff}(x) = \partial\ln \mathscr{R}(x,\sigma,1)/\partial \ln x$}.
The change in behaviour should occur around $x\approx 1$.
Then from (\ref{2.7}), the crossover scaling function can be read off
\BEQ
\mathscr{R}=\mathscr{R}(r,\sigma,\lambda) = \frac{\sigma r}{K_1(2 r^{1/2} \lambda^{-1})}
\int_0^{\infty} \!\D z\: \e^{-\sigma r^{1/2} z}\: K_0\left( \frac{2}{\lambda}\sqrt{ r(\lambda+1)\,}\,\right),
\EEQ
which is indeed invariant under the rescaling $r'=br$, $\sigma'=\sigma$, $\mu'=\mu$ and $\lambda'=b^{1/2}\lambda$. {Hence $\phi=2
$.}
Choosing $b=\lambda^{-2}$,
we can identify $\mathscr{R}\left( X,\sigma,1\right)$  with $X=r\lambda^{-2}$ such that the crossover scaling function reads
\BEQ\label{4.5}
\mathscr{R}\left(X,\sigma,1\right) = \frac{\sigma X}{K_1(2 X^{1/2})} \int_0^{\infty}\!\D z\: e^{-\sigma X^{1/2} z}\: K_0\left(2\sqrt{X(z+1)\,}\,\right).
\EEQ

\begin{figure}[tb]
\begin{center}
\includegraphics[width=.5\hsize]{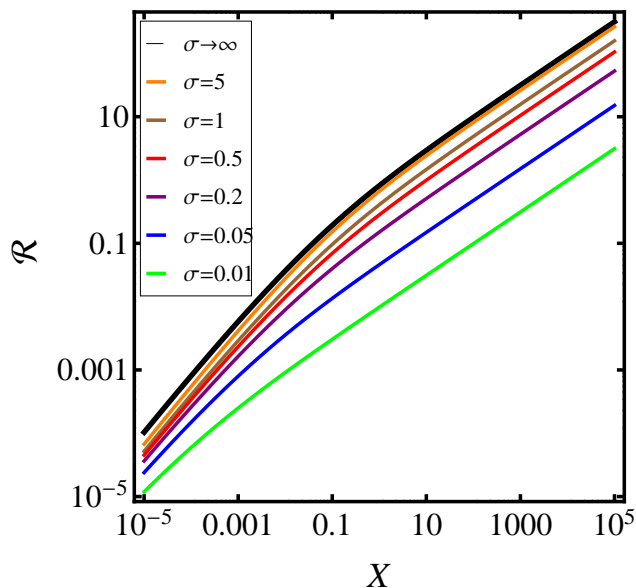}
\end{center}
\caption[fig5]{Crossover scaling function $\mathscr{R}=\mathscr{R}(X,\sigma,1)$
{of the stationary state of the coagulation-diffusion process with a stochastic reset}, for several initial dimensionless concentrations $\sigma$.
The thick black line is obtained from eq.~(\protect\ref{4.7}).
}
\label{fig5}
\end{figure}

In figure~\ref{fig5}, {the dependence of this scaling function on the scaling variable $X$ is shown} for several initial concentrations $\sigma$.
The change in behaviour from $\beta_{\rm MF}=1$ on the Bethe lattice (for $X\ll 1$) to the value $\beta=\demi$
on the chain (with $X\gg 1$) is {depicted}. We also find that
the crossover is more clearly defined for larger initial densities. This appears natural since for small densities the lattice is only sparsely
populated and correlation effects become less pronounced.

\begin{figure}[tb]
\begin{center}
\includegraphics[width=.5\hsize]{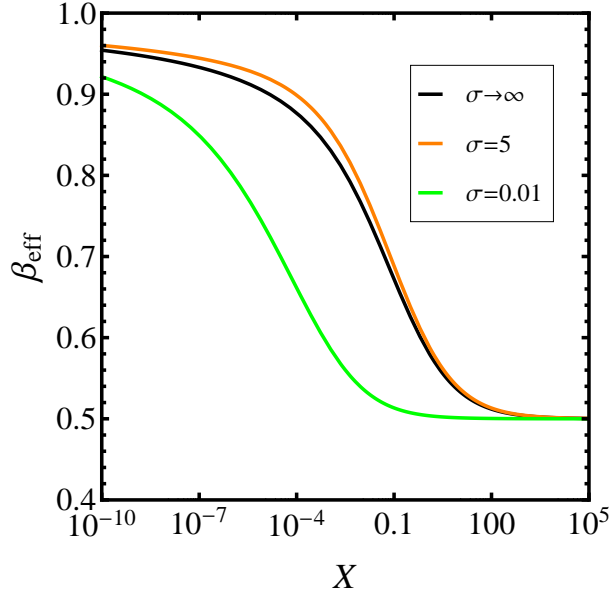}
\end{center}
\caption[fig4]{Dependence of effective exponent $\beta_{\rm eff}$ on $X$
{of the stationary state of the coagulation-diffusion process with a stochastic reset},
for \textcolor{black}{three} initial dimensionless concentrations $\sigma$.
The line $\sigma\to\infty$ is the logarithmic derivative of (\protect\ref{4.7}).
}
\label{fig4}
\end{figure}

The limit $\sigma\to\infty$ can be {determined explicitly}: changing variables $z\mapsto z \sigma $ in eq.~(\ref{4.5}), we have
\BEQ \label{4.6}
\mathscr{R}\left(X,\sigma,1\right) = \frac{ X}{K_1(2 X^{1/2})} \int_0^{\infty}\!\D z\: e^{- X^{1/2} z}\: K_0\left(2\sqrt{X(z/\sigma+1)\,}\,\right).
\EEQ
Taking the limit $\sigma\to\infty$, with $X$ fixed, the term $z/\sigma$ in the argument of $K_0$ can be neglected, hence
\BEQ\label{4.7}
\mathscr{R_\infty}(X)=\lim_{\sigma\to\infty} \mathscr{R}\left(X,\sigma,1\right) = \frac{ X^{1/2} K_0(2 X^{1/2})}{K_1(2 X^{1/2})} .
\EEQ
This leads to the asymptotics
\BEQ \label{4.8}
\mathscr{R_\infty}(X)\simeq \left\{ \begin{array}{ll} X^{1/2}                              & \mbox{\rm ~;~~ if $X\to\infty$}, \\
                                                 X\left[\ln X^{-1} \: -2C_E \right]   & \mbox{\rm ~;~~ if $X\to 0$}.
                                    \end{array} \right.
\EEQ
Figure~\ref{fig4} shows the behaviour of the effective exponent  $\beta_{\rm eff}$ as function of scaling variable $X$,
for three initial concentrations $\sigma$. While for large values of $X$, $\beta_{\rm eff}$ is close to $\demi$, as expected, its value approaches
unity as $X\to 0$. In contrast to crossover behaviour described by pure power-laws \cite{Lueb04,Henkel08},
this approach is logarithmically slow in the case at hand, see (\ref{4.8}). In addition, the crossover becomes more sharp with increasing $\sigma$.

A 
similar treatment can be carried out for the time-dependence of the particle-density. Comparing the equations of motion (\ref{1.7},\ref{3.1}),
we see that $r$ and $s$ play analogous roles. From (\ref{3.15}), we have the scaling form of the time-integrated density
$\lap{P}(s) s^2 = w \mathscr{P}(u,w)$, with the scaling function (\ref{2.7}) and
the modified definition $u=\sqrt{s\mu}/\lambda$ of the first scaling variable while $w=c\mu/\lambda$ is kept {fixed}.
This implies that the exponent $\beta$ of the stochastic reset (viz. $\varrho\sim r^{\beta}$) and $\alpha$ of the density decay exponent
(viz. $\varrho(t)\sim t^{-\alpha}$) are related through
\BEQ
\beta = \alpha.
\EEQ
In particular, if we relabel $\beta_{\rm eff} \mapsto  \alpha_{\rm eff}$ in figure~\ref{fig4}, we also have the crossover of the effective decay
exponent $\alpha_{\rm eff}$ between the chain for $X\gg 1$ and the Bethe lattice for $X\ll 1$.

\section{Conclusions}

{
We have studied the crossover of the coagulation-diffusion-process between the chain and the Bethe lattice
as an analytically treatable example of the crossover between diffusion-limited and reaction-limited kinetic reactions.
Using the ben Avraham-Glasser approximation, the empty-interval method can be extended from the chain (where it is exact) to the
Bethe lattice. Therein, the number $q$ of branches of the Bethe lattice is treated as a continuous variable. Taking the continuum limit $a\to 0$
also requires to let simultaneously $q\to 2$, see (\ref{1.4}), such that the {\it a priori} uncontrolled ben Avraham-Glasser
approximation becomes exact in this continuum limit. The model's behaviour
is then described in terms of two scaled couplings $\lambda,\mu$, where $\mu>0$ is a fixed constant while changing $\lambda$ from zero to
infinity describes the crossover from the chain to the Bethe lattice.}

{It turned out that a simple scaling analysis of the model's equation of motion does not work. We have seen that this comes from unexpected
logarithmic contributions to the universal long-time behaviour which also affects the crossover scaling between the chain and the Bethe lattice,
see figure~\ref{fig4}.
Such a logarithmic behaviour is unexpected from a standard mean-field treatment,
whereas the latter one might have anticipated from the fact that the Bethe lattice is infinite-dimensional.}

{For the understanding of this crossover, the analysis of the non-equilibrium stationary state,
reached by the coagulation-diffusion process subject to a stochastic reset, has proven to be a valuable tool, but is also of interest in its own right.
We have seen that on the Bethe lattice, the near-critical behaviour obtained for small reset rates $r\to 0$
displays logarithmic corrections to scaling as one would expect for a system at its upper critical dimension.
Using a mathematical duality between the reset rate $r$ for the stationary state with a reset and the Laplace variable $s$ conjugate to the time $t$
for the time-dependence of the non-stationary model without a reset,
it is necessary to study first the time-integrated particle-density $P(t)$ which has indeed a logarithmic long-time decay.
In contrast to the expectation from mean-field theory, the model's behaviour on the Bethe lattice is surprisingly characterised by additional
logarithmic factors. No larger deviation from a simple mean-behaviour is possible for a system at high spatial dimensions. This suggests that
{\it the coagulation-diffusion process on the Bethe lattice should behave rather like a system at its upper critical dimension}, instead of having
the behaviour of a system at infinite spatial dimensionality. This feature did not appear in previous studies of the equilibrium critical behaviour of
statistical systems on the Bethe lattice.}

{The only quantity which does not seem to fit into this picture is the time-dependent density $\varrho(t)=\partial_t P(t)$. However, in this
case taking the derivative converts the behaviour of the slowly varying observable $P(t)$ to an algebraically rapid decay of $\varrho(t)$.
It is mathematically well-established that the behaviour of a derivative $\partial_t P(t)$ can be considerably more irregular than that of a function
$P(t)$ itself \cite{Feller71}.}

{Our precise analysis also leads to the dynamical exponent $z=2$, in agreement with the microscopic diffusive motion of the individual single particles.
This result contradicts earlier assertions that $z=1$, which would have suggested ballistic transport of single particles, and which were based on the
unjustified simplification of setting $\mu=0$ in the equation of motion (\ref{1.5}).
}

It would be of interest to compare the explicitly known crossover scaling functions of $\varrho(t)$ and $P(t)$ with experimental data \cite{Allam13}.\\~\\

\noindent
{\bf Acknowledgements:}
MD and XD
acknowledge the hospitality of the Groupe de Physique Statistique (now at LPCT) during their stay at
the Universit\'{ e} de Lorraine Nancy, where most of the work on
this project has been performed.  MD  and DS are supported  by FP7 EU IRSES
projects No. 612707 ``Dynamics of and in Complex Systems''. XD is supported by the National Research
Foundation of Korea (NRF) grant funded by the Korea
government (MSIP) (No. 2016R1A2B2013972).



\newpage
{\small\begin{thebibliography}{999}
\bibitem{Abra65} M. Abramowitz and I.A. Stegun, {\it Handbook of Mathematical Functions}, Dover (New York 1965).
\bibitem{Abad00} E. Abad, H.L. Frisch and G. Nicolis, J. Stat. Phys. {\bf 99}, 1397 (2000) {\tt [arXiv:cond-mat/0002432]}.
\bibitem{Abad02} E. Abad, T. Masser and D. ben Avraham, J. Phys. A Math. Gen. {\bf 35}, 1483 (2002) {\tt [arXiv:cond-mat/0201446]}.
\bibitem{Abad04} E. Abad, Phys. Rev. {\bf E70}, 031110 (2004) {\tt [arxiv:cond-mat/0501245]}.
\bibitem{Agha05} A. Aghamohammadi and M. Khorrami, Eur. Phys. J. {\bf B47}, 583 (2005) {\tt [cond-mat/0511649]}.
\bibitem{Ali09} M. Alimohammadi and N. Olanj, Physica {\bf A389}, 1549 (2010) {\tt [arXiv:0904.0847]}.
\bibitem{Allam13} J. Allam, M.T. Sajjad, R. Sutton, K. Litvinenko, Z. Wang, S. Siddique, Q.-H. Yang, W.H. Loh and T. Brown,
  Phys. Rev. Lett. {\bf 111}, 197401 (2013) {\tt [arxiv:1310.4437]}.
\bibitem{Baxter82} R.J. Baxter, {\it Exactly solved models in statistical mechanics}, Academic Press (London 1982).
\bibitem{benA90} D. ben Avraham, M. Burschka and C.R. Doering, J. Stat. Phys. {\bf 60}, 695 (1990).
\bibitem{benA00} D. ben Avraham and S. Havlin, {\it Diffusion and Reactions in Fractals and Disordered Systems},
                 Cambridge University Press (Cambridge 2000).
\bibitem{benA06} D. ben Avraham and M.L. Glasser, J. Phys. Condens. Matter {\bf 19}, 065107 (2006) {\tt [arXiv:cond-mat/0612089]}.
\bibitem{Chatelain12} C. Chatelain, M. Henkel, M. J. de Oliveira and T. Tom\'{e}, J.  Stat. Mech. P11006 (2012) {\tt [arxiv:1207.2247]}.
\bibitem{Cadilhe04} A. Cadilhe and V. Privman, Mod. Phys. Lett. {\bf B18}, 207 (2004) {\tt [arXiv:cond-mat/0311190]}.
\bibitem{Cadilhe07} A. Cadilhe, N.A.M. Ara\'{u}jo and V. Privman, J. Phys.: Condens. Matter {\bf 19}, 065124 (2007) {\tt [arXiv:cond-mat/0611413]}.
\bibitem{Dahmen95} S.R. Dahmen, J. Phys. A: Math. Gen. {\bf 28}, 905 (1995) {\tt [cond-mat/9405031]}.
\bibitem{Dudka17} M. Dudka, O. B\'{e}nichou and G. Oshanin, J. Stat. Mech.  043206 (2018) {\tt [arXiv:1710.07934]}.
\bibitem{Durang10} X. Durang, J.-Y. Fortin, D. del Biondo, M. Henkel and J. Richert, J. Stat. Mech. P04002 (2010) {\tt \mbox{\tt[arxiv:1012.4724]}}.
\bibitem{Durang11} X. Durang, J.-Y. Fortin and M. Henkel, J. Stat. Mech. P02030 (2011) {\tt [arxiv:1001.3526]}.
\bibitem{Durang14} X. Durang, M. Henkel and H. Park, J. Phys. A Math. Theor. {\bf 47}, 045002 (2014) \mbox{\tt [arxiv:1309.2107]}.
\bibitem{Evans11a} M.R. Evans and S.N. Majumdar, Phys. Rev. Lett. {\bf 106}, 160601 (2011) {\tt [arxiv:1102.2704]}.
\bibitem{Evans11b} M.R. Evans and S.N. Majumdar, J. Phys. A Math. Theor. {\bf 44}, 435001 (2011) {\tt [arxiv:1107.4225]}.
\bibitem{Evans14} M.R. Evans and S.N. Majumdar, J. Phys. A Math. Theor. {\bf 47}, 285001 (2014) {\tt [arxiv:1404.4574]}.
\bibitem{Falcao17} R. Falcao and M.R. Evans, J. Stat. Mech. P023204 (2017) {\tt [arxiv:1610.03503]}.
\bibitem{Feller71} W. Feller, {\it An introduction to probability theory and its applications}, vol. 2 (2$^{\rm nd}$ ed), Wiley (New York 1971).
\bibitem{Fortin14} J.-Y. Fortin, J. Stat. Mech. P09033 (2014) {\tt [arxiv:1402.6901]}.
\bibitem{Fortin17} J.-Y. Fortin, {\tt [arxiv:1711.03703]}.
\bibitem{Fuchs16} J. Fuchs, S. Goldt and U. Seifert, Europhys. Lett. {\bf 113}, 60009 (2016) {\tt [arxiv:1603.01141]}.
\bibitem{Henkel96} M. Henkel and F. Seno, Phys. Rev. {\bf E53}, 3662 (1996) {\tt [arxiv:cond-mat/9601105]}.
\bibitem{Henkel01} M. Henkel and H. Hinrichsen, J. Phys. A: Math. Gen. {\bf 34}, 1561 (2001) {\tt [cond-mat/0010062]}.
\bibitem{Henkel08} M. Henkel, H. Hinrichsen and S. L\"ubeck,
  {\it ``Non-equilibrium phase transitions vol. 1: absorbing phase transitions''}, Springer (Heidelberg 2008).
\bibitem{Igloi88} F. Igl\'oi and L. Turban, Phys. Rev. {\bf E78},  031128 (2008)  {\tt [arxiv:0808.3512]}
\bibitem{Kamke77} E. Kamke, {\it ``Differentialgleichungen: L\"osungsmethoden und L\"osungen I''}, Teubner (Stuttgart 1997).
\bibitem{Kapr10} P.L. Kaprivsky, S. Redner and E. Ben-Naim, {\it ``A kinetic view of statistical physics''},
  Cambridge University Press (Cambridge 2010).
\bibitem{Khorr03} M. Khorrami, A. Aghamohammadi and M. Alimohammadi, J. Phys. A: Math. Gen. {\bf 36}, 345 (2003) {\tt [cond-mat/0112490]}.
\bibitem{Khorr14} M. Khorrami and A. Aghamohammadi, J. Stat. Mech. P07017 (2014) {\tt [arXiv:1407.5591]}.
\bibitem{Kopelman90} R. Kopelman, C.S. Li and Z.-Y. Shi, J. Lumin. {\bf 45}, 40 (1990).
\bibitem{Krebs95} K. Krebs, M.P. Pfannm\"uller, B. Wehefritz and H. Hinrichsen, J. Stat. Phys. {\bf 78}, 1429 (1995)
  {\tt [cond-mat/9402017], [cond-mat/9402018], [cond-mat/9402019]}.
\bibitem{Kroon93} R. Kroon, H. Fleurent and R. Sprik, Phys. Rev. {\bf E47}, 2462 (1993).
\bibitem{Lueb04} S. L\"ubeck, Int. J. Mod. Phys. {\bf b18}, 3977 (2004) {\tt [arXiv:cond-mat/0501259]}.
\bibitem{Masser00} T. Masser and D. ben-Avraham, Phys. Lett. {\bf A275}, 382 (2000). {\tt [cond-mat/0008448]}.
\bibitem{Majumdar93} S. N. Majumdar, V. Privman, J. Phys. A Math. Gen. {\bf 26}, L743 (1993) {\tt [arxiv:cond-mat/9305030]}.
\bibitem{Matin07} L.F. Matin, A. Aghamohammadi, and M. Khorrami, Eur. Phys. J. {\bf B56}, 243 (2007).
\bibitem{Matin15} L.F. Matin, J. Theor. Appl. Phys. {\bf 9}, 93 (2015).
\bibitem{Matt98} D.C. Mattis and M.L. Glasser, Rev. Mod. Phys. {\bf 70}, 979 (1998).
\bibitem{Mazilu12} D.A. Mazilu, I. Mazilu, A.M. Seredinski, V. O. Kim, B. M. Simpson and W. E. Banks,
  J. Stat. Mech. P09002 (2012) {\tt [arxiv:1207.2168]}.
\bibitem{Montero17} M. Montero, A. Mas\'o-Puigdellosas and J. Villaroel, Eur. Phys. J. {\bf B90}, 176 (2017) {\tt [arXiv:1706.04812]}.
\bibitem{Muna06a} R. Munasinghe, R. Rajesh, R. Tribe and O. Zaboronski, Comm. Math. Phys. {\bf 268}, 717 (2006).
\bibitem{Muna06b} R. Munasinghe, R. Rajesh and O. Zaboronski, Phys. Rev. {\bf E73}, 051103 (2006) {\tt [cond-mat/0506398]}.
\bibitem{Mura09} Y. Murakami and J. Kono, Phys. Rev. Lett. {\bf 102}, 037401 (2009).
\bibitem{LeVot18} F. Le Vot, C. Escudero, E. Abad, S.B. Yuste, {\tt  [arXiv:1804.03213]}.
\bibitem{Ostilli12} M. Ostilli, Physica {\bf A391}, 3417 (2012) {\tt [arXiv:1109.6725]}.
\bibitem{Pras89} J. Prasad and R. Kopelman, Chem. Phys. Lett. {\bf 157}, 535 (1989).
\bibitem{Prud1} A.P. Prudnikov, Yu.A. Brychkov, O.I. Marichev, {\it Integrals and series, vol. 1: elementary functions},
  Gordon \& Breach (New York  1986).
\bibitem{Prud4} A.P. Prudnikov, Yu.A. Brychkov, O.I. Marichev, {\it Integrals and series, vol. 4: direct Laplace transforms},
  Gordon \& Breach (New York  1992).
\bibitem{Prud5} A.P. Prudnikov, Yu.A. Brychkov, O.I. Marichev, {\it Integrals and series, vol. 5: inverse Laplace transforms},
  Gordon \& Breach (New York  1992).
\bibitem{Racz85} Z. R\'acz, Phys. Rev. Lett. {\bf 55}, 1707 (1985).
\bibitem{Rey97} P.-A. Rey and M. Droz, J. Phys. A: Math. Gen. {\bf 30}, 1101 (1997) {\tt [cond-mat/9609088]}.
\bibitem{Rios01} P. de los Rios, S. Lise and A. Pelizzola, Europhys. Lett. {\bf 53}, 176 (2001) {\tt [arxiv:cond-mat/0011258]}.
\bibitem{Roldan16} \'E. Rold\'an, A. Lisica, D. S\'anchez-Taltavull and S.W. Grill,
  Phys. Rev. {\bf E93}, 062411 (2016) {\tt [arxiv:1603.06956]}.
\bibitem{Roldan17} \'E. Rold\'an and S. Gupta, Phys. Rev. {\bf E96}, 022130 (2017) {\tt [arXiv:1703.10615]}.
\bibitem{Rozikov08} U.A. Rozikov, J. Stat. Phys. {\bf 130}, 801 (2008)  {\tt [arxiv:math-ph/0611038]}.
\bibitem{Russo06} R.M. Russo, E.J. Mele, C.L. Kane, I.V. Rubtsov, M.J. Therien and D.E. Luzzi,
  Phys. Rev. {\bf B74}, 041405(R) (2006).
\bibitem{Spouge88} J.L. Spouge, Phys. Rev. Lett. {\bf 60}, 871 (1988); erratum Phys. Rev. Lett. {\bf 60}, 1885 (1988).
\bibitem{Sriv09} A. Srivastava and J. Kono, Phys. Rev. {\bf B79}, 205407 (2009).
\bibitem{Tous83} D. Toussaint and F. Wilczek, J. Chem. Phys. {\bf 78}, 2642 (1983).
\bibitem{Turban80} L. Turban, Phys. Lett. {\bf 78A}, 404 (1980).
\bibitem{Turban17} L. Turban and J.-Y. Fortin, {\tt [arxiv:1711.01248]}.
\bibitem{Yuste01} S.B. Yuste and K. Lindenberg, Phys. Rev. Lett. {\bf 87}, 118301 (2001) {\tt [arXiv:cond-mat/0105338]}.
\bibitem{Yuste02} S.B. Yuste and K. Lindenberg, Chem. Phys. {\bf 284}, 169 (2002).
\end{thebibliography}
}
\end{document}